# Nematic superconductivity from selective orbital pairing in Ba(Fe$_{1-x}$M$_x$)$_2$As$_2$ (M = Co, Ni) single crystals

Mason Klemm[1], Shirin Mozaffari[2], Rui Zhang[1], Brian W. Casas[2], Alexei E. Koshelev[3], Ming Yi[1], Luis Balicas[2,*], and Pengcheng Dai[1,**]

1. Department of Physics and Astronomy, Rice University, Houston, TX 77005 USA
2. National High Magnetic Field Laboratory, Florida State University, Tallahassee, FL 32310 USA
3. Materials Science Division, Argonne National Laboratory, Lemont, IL 60439 USA

**We use transport measurements to determine the in-plane anisotropy of the upper critical field $H_{c2}$ in detwinned superconducting Ba(Fe$_{1-x}$M$_x$)$_2$As$_2$ (M = Co, Ni) single crystals. In previous measurements on twinned single crystals, the charge carrier doping dependence (x) of the upper critical field anisotropy for fields along the inter-planar (*c*-axis) and in-plane field directions was found to increase in the overdoped regime. For underdoped samples, which exhibit a spin nematic phase below the tetragonal to orthorhombic structural transition temperature $T_s$, we find that $H_{c2}$ along the *a*-axis is considerably lower than that along the *b*-axis. The upper critical field anisotropy disappears in the over-doped regime when the system becomes tetragonal. By combining these results with inelastic neutron scattering studies of spin excitations, and angle-resolved photoemission spectroscopy, we conclude that superconductivity in under-doped iron pnictides is orbital selective – with a dominant contribution from electrons with the $d_{yz}$ orbital character and being intimately associated with spin excitations.**

In conventional Bardeen–Cooper–Schrieffer (BCS) superconductors, electrons form coherent Cooper pairs below the superconducting transition temperature through interacting with the lattice [1]. If the electrons forming Cooper pairs mostly originate from a single electronic band, the superconducting state is the ground state and does not coexist with charge- or spin- density wave order originating from the superconducting electronic band [2]. However, there are many multiband BCS superconductors, such as $MgB_2$ [3,4] and $NbSe_2$ [5,6], where superconductivity can occur in a specific band with an orbital selective superconducting gap. While unconventional copper oxide superconductors are single band materials [7-9], iron-based superconductors are known multiband materials where superconductivity may occur through an orbital-selective pairing mechanism analogous to that of multiband BCS superconductors [10-13]. Most unconventional superconductors, in contrast to BCS superconductors, have nonsuperconducting antiferromagnetic (AF) ordered parents and superconductivity arises from electron/hole doping that suppresses the AF order [7-9,14,15]. In the case of iron pnictide superconductors, there is an additional tetragonal-to-orthorhombic structural phase transition at $T_s$, occurring at or above the collinear static AF ordering temperature $T_N$ [16]. The structural transition is associated with an electronic nematic phase, where transport and electronic properties exhibit twofold rotational symmetry within the nearly square FeAs lattice [17-21], and orbital dependent Fermi surface nesting [22-28]. With increasing electron doping, $T_s$ and $T_N$ decrease and superconductivity appears as shown schematically in Fig. 1(a) for the $Ba(Fe_{1-x}M_x)_2As_2$ (M = Co, Ni) families of iron pnictide superconductors [29-33]. Given the close connection between magnetism and superconductivity, pairing in these superconducting materials is believed to be mediated by AF spin fluctuations [14,15], arising from quasiparticle excitations between the sign reversed hole and electron Fermi surfaces at different parts of the Brillouin zone

[34]. This manifests experimentally as a resonance, or a collective magnetic excitation at an energy $E_r$ appearing below the superconducting transition temperature $T_c$ with its intensity tracking the superconducting order parameter, in the spin excitation spectra [14,15,35-37]. In the weak-coupling theory of superconductivity, the resonance is a bound state arising from quasiparticle excitations that connect parts of the Fermi surfaces exhibiting a sign change in the superconducting order parameter $\Delta(k) = -\Delta(k+Q)$, where $\Delta(k)$ is the momentum ($k$)-dependent superconducting gap and $Q$ is the momentum transfer connecting the two gapped Fermi surfaces [14,15,35-37].

The left inset in Figure 1(a) shows the collinear static AF structure of $BaFe_2As_2$, the parent compound of $Ba(Fe_{1-x}M_x)_2As_2$ families of iron pnictide superconductors [15]. In the underdoped regime, these materials exhibit a tetragonal-to-orthorhombic structural transition at $T_s$, a collinear static AF order below $T_N$, and superconductivity below $T_c$ [$T_s > T_N > T_c$]. When the doping reaches the optimal level for superconductivity [x = 0.06 (M=Co) and 0.04 (M=Ni)] and beyond, the system becomes tetragonal at all temperatures. With further doping increase, superconductivity vanishes for doping levels above x = 0.13 (M=Co) [Fig. 1(a)] [29-33]. For the iron pnictides, angle resolved photoemission spectroscopy experiments reveal that three Fe $t_{2g}$ orbitals ($d_{xz}$, $d_{yz}$, $d_{xy}$) are active near the Fermi level forming hole and electron Fermi pockets as shown in Fig. 1(b) [13,38]. Nesting among electron-hole Fermi surface sheets associated with quasiparticles on the $d_{xz}$ and $d_{yz}$ orbitals occurs along the $Q_1$ and $Q_2$ directions, respectively, and can induce spin excitations along these two directions [39]. In the paramagnetic tetragonal state, the degenerate $d_{xz}$ and $d_{yz}$ orbitals induce these spin excitations with equal intensity at $Q_1$ and $Q_2$. In the underdoped regime upon cooling below $T_s$ the $d_{yz}$ band of the electron Fermi surface at X/Y increases in energy, while the $d_{xz}$ band decreases in energy [40,41]. Better nesting conditions resulting from the shifting of the bands lead to orbital selective spin excitations for $d_{yz}$ orbitals along the $Q_1$ direction [42,43]. If the neutron spin resonance is associated with the superconducting electron pairing, these results would suggest that superconductivity in iron arsenides and selenides is orbital selective [11,12,42,44,45] and dominated by the $d_{yz}$ orbitals with associated AF fluctuations along the $Q_1$ direction. For underdoped Ba(Fe$_{1-x}$M$_x$)$_2$As$_2$ where superconductivity co-exists microscopically with static AF order [15], there is a reduction of static ordered magnetic moment below $T_c$ due to its competition with superconductivity [29,46,47]. Therefore, one would expect an anisotropic superconducting state within the orthorhombic lattice plane of the underdoped regime that disappears when the crystal becomes tetragonal in the overdoped regime without AF order at all temperatures. Here, using transport measurements, we report that the anisotropy of the in-plane upper critical fields $H_{c2}$ of underdoped Ba(Fe$_{1-x}$M$_x$)$_2$As$_2$ is consistent with an orbital selective pairing that involves electrons having $d_{yz}$ orbital character.

The presence of twin domains in large single crystals of underdoped Ba(Fe$_{1-x}$M$_x$)$_2$As$_2$ renders conventional measurements insufficient to distinguish the rotational symmetry breaking between the $d_{xz}$ and $d_{yz}$ orbitals that occurs below $T_s$, meaning that neither orbital can be unambiguously determined to be associated with superconductivity in such a state. Therefore, uniaxial pressure must be used to obtain a single domain from which the intrinsic magnetic and electronic properties can be determined [48]. Electronic nematicity below $T_s$ has long been established in the Ba(Fe$_{1-x}$M$_x$)$_2$As$_2$ materials through the process of detwinning [17-21]. The schematic provided in the left inset of Fig. 1(a) details the in-plane AF ordering with respect to the direction of the applied uniaxial pressure. Recent neutron scattering experiments on detwinned, underdoped Ba(Fe$_{1-x}$Co$_x$)$_2$As$_2$ reveal anisotropic spin fluctuations, with the neutron spin resonance associated with superconductivity observed along the AF wavevector, $Q_1 = (1,0)$, and no magnetic scattering observed along $Q_2 = (0,1)$ [42]. Additionally, the Fermi surface is modified below $T_s$ in underdoped single crystals and, consequently, of the nesting condition [Figs. 1(b-c)] [13,40]. Although the upper critical field anisotropy for in-plane and *c*-axis aligned magnetic fields has been heavily investigated for different families of iron-based superconductors [33,49-56], there is no report of in-plane upper critical field anisotropy for Ba(Fe$_{1-x}$M$_x$)$_2$As$_2$ having the orthorhombic lattice structure in the underdoped regime.

Single-crystals were grown with Fe$_{1-x}$M$_x$As as the chosen flux via the self-flux method [57]. We chose to study underdoped Ba(Fe$_{1-x}$Ni$_x$)$_2$As$_2$ with x ~ 0.04, which has a $T_N$ ~44 K where AF order coexists with superconductivity below $T_c$ ~ 17 K [47,58], and overdoped Ba(Fe$_{1-x}$Co$_x$)$_2$As$_2$ with x = 0.11 [30]. Larger single crystals were aligned and later cut along the orthorhombic *a* and *b* axes into roughly 2x2 mm squares using a Laue x-ray diffractometer and a wire saw respectively. Samples were then detwinned via uniaxial pressure with a small brass clamp. To investigate a potential orbital selective pairing mechanism, we probe the anisotropy of the in-plane upper critical fields of underdoped Ba(Fe$_{1-x}$Ni$_x$)$_2$As$_2$ via resistivity measurements along the *c* axis by using a Corbino geometry and an in-plane DC field [Fig. 1(d)] [50]. We use a Corbino-like geometry for the electrical contacts to maximize the cross-sectional area of the sample within the *ab*-plane, and to ensure a higher homogeneity for the electrical current density flowing along the *c*-axis. High current homogeneity and large cross-sectional areas are not achievable with other configurations for the electrical contacts. Field dependent measurements of $T_c$ were first conducted using the rotator option of a Quantum Design Physical Properties Measurement System with the later measurements performed at the National High Magnetic Field Laboratory using both a superconducting 18 T magnet and a 41 T resistive Bitter magnet, respectively. Our results reveal a clear anisotropy in the upper critical fields of underdoped Ba(Fe$_{1-x}$Ni$_x$)$_2$As$_2$, with the field along the *a*-axis resulting in a lower projected $H_{c2}$ relative to fields aligned along the *b* axis (Figs. 2-4).

To systematically characterize $T_c$, a range of magnetic fields were applied along the *a* and *b* axes of each sample and then temperature was swept across the superconducting transition while maintaining a constant applied field. A schematic of the geometric relationship between the applied field direction, crystal axes, and strain direction is indicated in Fig. 1(d). It is clear from the raw resistivity data that a planar anisotropy exists in the underdoped samples [Fig. 2(a,c,e)]. Traces of identical colors correspond to the same magnitude of the applied magnetic field with the lighter and darker shade of the same color representing the direction of the field, i.e., along the *a* and *b* axes, respectively. From these temperature sweeps, we extract the onset of $T_c$ which we define as 90% of the value of the resistivity in the normal state just above $T_c$. A sample-dependent, anomalous hump appears before the onset of superconductivity in some of our measurements. Strain free measurements reveal the hump only appears after pressure is applied to the sample (see supplementary Fig. S1). In these cases, we treat the 90% criterion as 90% of the normal state resistivity before the upturn of the humps [Fig. 2(a)]. Note that the apparent anisotropy depends on the definition of the onset as reflected in a similar work [53]. Since the evaluation of the superconducting gap function would require an array of thermodynamic, spectroscopic, and transport techniques, transport measurements alone cannot conclusively determine the absolute magnitude of the superconducting gap anisotropy. Nevertheless, from the transport measurements we can evaluate the anisotropy of the characteristic energy scale required for thermally activated flux flow of vortices (see Supplemental Fig. S5).

One important question to address is the impact of magnetic order and associated magnetoresistance to the observed $H_{c2}$ anisotropy relative to fields aligned along the *a* and *b* axis. From in-plane magnetic field neutron diffraction measurements on twinned Ba(Fe$_{1-x}$Ni$_x$)$_2$As$_2$ with x ~ 0.04 ($T_c$ = 17 K and $T_N$ ≈ 44 K), an identical doping level as our current transport measurements, an in-plane magnetic field of 10-T has no observable impact on magnetic ordered moment, $T_N$, or spin fluctuations in the normal state (above $T_c$) but can recover superconductivity-suppressed static magnetic ordered moment below $T_c$ [47]. Since the magnetoresistance in the normal (nonsuperconducting) state of Ba(Fe$_{1-x}$Ni$_x$)$_2$As$_2$ with x = 0.04 ($T$ ~ 18 K) is at most 0.01% for a 20-T in-plane field and has no observable anisotropy for fields aligned along the *a* and *b* direction (Supplementary Figs. S6-7), we conclude that the observed $H_{c2}$ anisotropy for fields aligned along the *a* and *b* in Fig. 2 is due to the suppression of superconductivity and unrelated to the static AF order.

Because the electronic nematic phase is believed to be associated with the unidirectional spin resonance observed by inelastic neutron scattering, and thus the orbital selective pairing, we can explore the overdoped side of the phase-diagram to further test this hypothesis [17,30,42,52]. Overdoped single crystals experience no structural phase transitions associated with electronic nematicity but do retain a superconducting transition at a comparable $T_c$ [Fig. 1(a)]. Identical resistivity measurements with uniaxial strain on overdoped samples reveal no anisotropy in the upper critical fields like the one seen for the underdoped single crystals [Fig. 2(b,d,f)]. Therefore, the anisotropy of $H_{c2}$ observed in the underdoped sample and the lack of thereof in the overdoped sample indicate that the observed anisotropic $H_{c2}$ in the former is not due to applied uniaxial pressure, but to the orbital selective nature of the pairing contingent upon the existence of an electronic nematic phase.

The Montgomery method was utilized to capture the planar resistivity when the magnetic field is applied along either the *a* or *b* axes as pictured in the inset of Figs. 3(b,d) [59,60]. The purpose of these measurements is to test the differences in resistivity for current along and perpendicular to the applied magnetic field direction. Figures 3(a) and 3(c) summarize the results for underdoped single crystals with current along the *a*-axis and *b*-axis, respectively. While fields along the *b*-axis suppress $T_c$ more significantly than fields along the *a*-axis according to inter-planar resistivity measurements, the suppression of $T_c$ as probed via the planar resistivity depends on the relative orientation between the field and the electrical currents. Specifically, $T_c$ experiences a more dramatic suppression near the transition temperature when the applied field and the electric current are perpendicular to one another. A sample on the overdoped side with a slightly higher $T_c$, measured via this configuration, still reveals a very small anisotropy [Figs. 3(b-d)]. The higher $T_c$ of the overdoped samples is consistent with a closer proximity to the nematic phase, which vanishes near optimal superconductivity, compared with overdoped samples with the same $T_c$ as that of the underdoped one [Fig. 1(a)]. In this configuration, one cannot exclude the possibility that this anisotropy might result from the relative alignment between the electrical currents and the external field or their interactions. Nevertheless, temperature sweeps for various applied currents reveal negligible differences in $T_c$, ruling out the magnitude of the current as being responsible for the observed anisotropy (see Supplementary Fig. S2). Therefore, the *c*-axis resistivity is the most appropriate configuration to characterize this anisotropy because the relationship between the externally applied field and the electrical current is maintained constant, in other words they are always kept perpendicular to each other. In particularly, there are clear differences in zero resistance temperatures for field applied along the *a*-axis and *b*-axis in the underdoped sample for fields above 20-T, when $T_c$ is suppressed to below 10 K [Fig. 2(e)]. Since we do not have data above 20-T in the Montgomery method discussed in Fig. 3, it is difficult to reach a conclusion about in-plane upper critical field anisotropy using these results.

Further analysis on the raw data in Fig. 2 was performed to extract the evolution of the onset $T_c$ as a function of the increasing field for both underdoped and overdoped samples [Figs. 4(a-b)]. Clearly, the anisotropy in $H_{c2}$ exists exclusively in the underdoped sample in this representation. The conditionality of the planar $H_{c2}$ in the underdoped samples becomes more evident by plotting the $T_c$ onset, demonstrating a faster suppression of the transport $T_c$ near the zero-field transition temperature when the field and the electrical current are perpendicular [Figs. 4(c-d)]. Plotting $\rho_a$ and $\rho_b$ when the field and current are parallel reveals minor anisotropy that increases at higher field [Fig. 4(e)]. The case where the field and electrical current are perpendicular result in no anisotropy within error bars [Fig. 4(f)].

In general, the anisotropy of the upper critical field may be caused by either anisotropy of Fermi velocity or gap and the first factor is more common. For the case of underdoped Ba(Fe$_{1-x}$Ni$_x$)$_2$As$_2$ materials with nematic order, several factors may contribute to the anisotropy of $H_{c2}$ : (i) the hole bands become anisotropic, (ii) anisotropies and occupations of the two electron bands become different, and (iii) the interband pairings between the hole bands and two electron bands become different leading to different superconducting gaps in the two electron bands. We believe that the last mechanism likely dominates due to strong modification of the nesting conditions upon entering the nematic state. Upon cooling from the normal state, the $d_{yz}$ band along the Γ-X direction is lifted towards the Fermi energy ($E_F$) while the $d_{xz}$ band along Γ-Y drops in energy relative to $E_F$, thus breaking the degeneracy between the two bands [13]. This should favor nesting between the Fermi surface sheets associated with the $d_{yz}$ bands when compared to those of the $d_{xz}$ bands as illustrated in Figs. 1(b-c). Such a nesting condition would provide a clear explanation for the observed magnetic scattering along $Q_1$ and $Q_2$ in recent work [42,43]. If superconducting electron pairing in the iron pnictides is driven by spin fluctuations, as is widely suspected, we should expect pairing to occur as the result of $d_{yz}$-$d_{yz}$ intra-orbital fluctuations, *i.e.,* an orbital selective pairing mechanism. A dramatic consequence of such a pairing mechanism would mean that the superconducting gaps are different for electrons with $d_{yz}$ and $d_{xz}$ orbital characters along the (1,0) and (0,1) directions, respectively. Therefore, we would also expect the upper critical fields of this system, at which the superconducting pairing breaks down, to be directionally dependent between the (1,0) and (0,1) directions in the underdoped regime. Our results in Fig. 4 reflect this anisotropy. It follows, then, that the overdoped single crystals should see no evidence of anisotropy among in-plane upper critical fields, as these samples experience no transition into a nematic state and therefore no modification of the nesting condition in the band structure. Indeed, there is no upper critical field anisotropy for overdoped samples under a similar uniaxial strain [Fig. 4(b)]. Therefore, our systematic upper critical field measurements support the notion that superconductivity in the underdoped iron pnictide superconductors is orbital selective, or clearly different from an isotropic *s*-wave superconductors. These results are consistent with twofold oscillations of angular dependence of the in-plane and out-of-plane magnetoresistivity of $Ba_{0.5}K_{0.5}Fe_2As_2$ near $T_c$ when the direction of the applied field is rotated within the plane [61]. We note that similar anisotropic upper critical fields have been reported for twisted bilayer graphene, as a sign of nematic superconducting pairing [62]. In that case, orbital-selective superconductivity is also the proposed mechanism, involving multi-component pairing [63].

We note that several Fe based superconductors display upper critical fields exceeding the Pauli limiting field [49]. The reason for this remains controversial, although it was often ascribed to the multi-band nature of the superconductivity in these compounds. One possibility, particularly for more disordered chemically doped samples, is that this renormalization results from spin orbit scattering, usually treated in terms of an isotropic spin-orbit scattering time $\tau_{so}$ [64]. Alternatively, anisotropy of the potential scattering time may also contribute to the observed anisotropy.

It is well known that the scattering rate in the Ba(Fe$_{1-x}$M$_x$)$_2$As$_2$ family of Fe based pnictide superconductors is rather anisotropic [20,65,66]. The reason for this anisotropy is not well understood, but it might reflect the intrinsic nematicity of the underlying electronic fluid from which superconductivity condenses [67]. In this scenario, one would expect that $\tau_{so}$ becomes anisotropic in the underdoped Ba(Fe$_{1-x}$M$_x$)$_2$As$_2$ compounds below $T_N$ but above $T_c$, leading to the anisotropy in upper critical fields observed here. However, we find a very small resistivity anisotropy associated with the electronic nematicity in the underdoped region of these compounds (See supplementary Fig. S6).

In summary, we use a series of resistivity measurements to analyze the anisotropy of the in-plane $H_{c2}$s for uniaxial strain detwinned crystals of the iron pnictide superconductor Ba(Fe$_{1-x}$M$_x$)$_2$As$_2$. For underdoped samples with an orthorhombic lattice structure, we find a clear in-plane upper critical field anisotropy in uniaxial strain detwinned samples for fields above 20-T with a $c$-axis current. These results are consistent with an orbital selective electron pairing, where the superconducting gap along the (1,0) direction with $d_{yz}$ orbital character is larger than that along the (0,1) direction with $d_{xz}$ orbital character. Therefore, superconductivity in the underdoped regime acquires a "nematic" character when it condenses from this electronic/spin state [67]. For the overdoped samples with a tetragonal lattice structure and without a nematic phase, similar measurements show no evidence of anisotropy in the in-plane upper critical fields. These results are consistent with an orbital selective pairing mechanism, suggesting that the quasiparticle excitations between electron-hole Fermi surface pockets having a $d_{yz}$ orbital character are favorable for electron pairing. These results further confirm that superconductivity is likely mediated by spin fluctuations associated with the electron-hole quasiparticle excitations in iron based superconductors [14].

The experimental and basic materials synthesis work at Rice is supported by the U.S. DOE, BES under Grant No. DE-SC0012311 and by the Robert A. Welch Foundation under Grant No. C-1839, respectively (P.D.). L.B. is supported by the US DOE, Basic Energy Sciences program through award DE-SC0002613. M.Y. acknowledges support by the Gordon and Betty Moore Foundation's EPiQS Initiative through grant no. GBMF9470, and the Robert A. Welch Foundation, Grant No. C-2175. The National High Magnetic Field Laboratory acknowledges support from the US-NSF Cooperative agreement Grant number DMR-1644779 and the state of Florida. Work of A.E.K at Argonne National Laboratory was funded by the US Department of Energy, Office of Science, Basic Energy Sciences, Materials Sciences and Engineering Division.

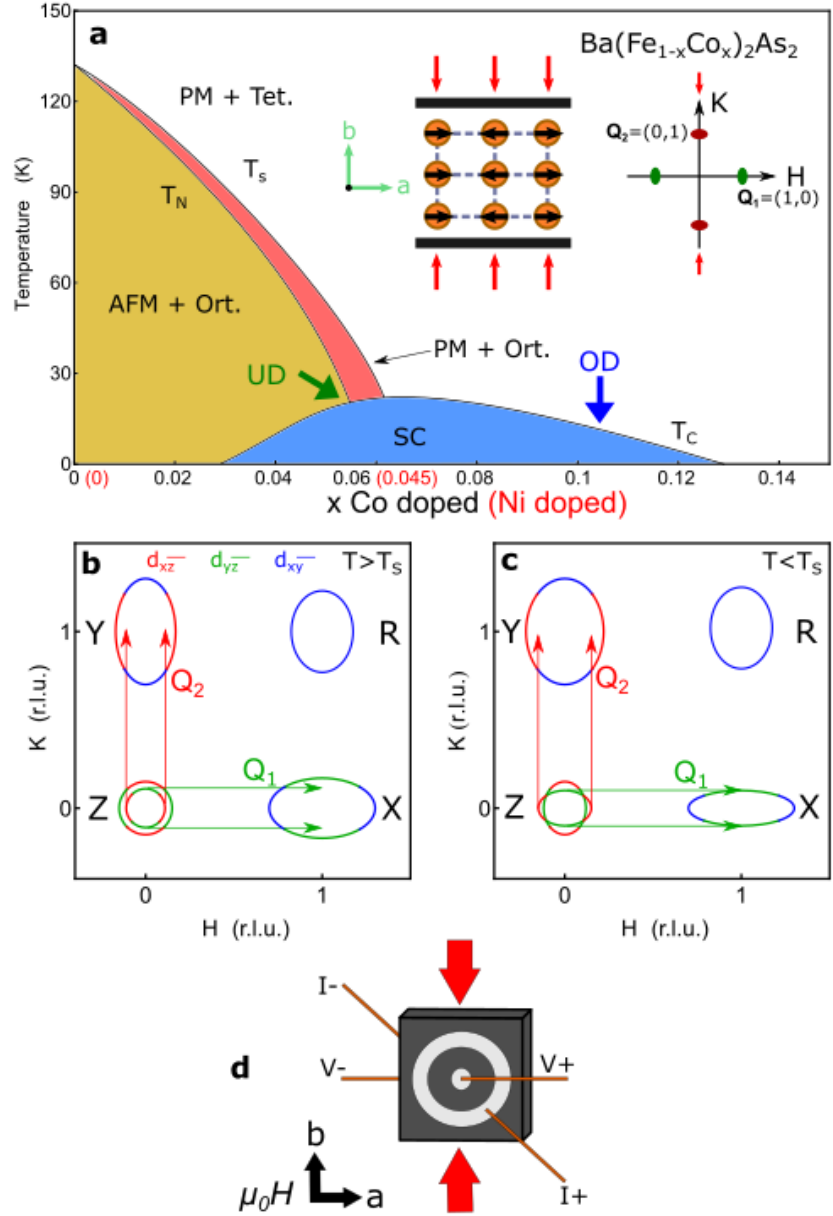

**Fig. 1** (a) Phase diagram denoting the rough doping of the underdoped (UD) and overdoped (OD) samples. Note that the horizontal axis is modified for Ni doped samples, as Ni contributes approximately twice as many electrons as Co. The first inset describes the orientation of the strain (red arrows) relative to the antiferromagnetic ordering defining the *a* and *b* axes. The second inset reflects the AF wavevector $Q_1$ (green) and wavevector $Q_2$ (red) studied in prior neutron scattering work [29-32]. (b) Illustrative Fermi surfaces and associated orbital contributions for the paramagnetic tetragonal lattice and the (c) nematic state below $T_s$ for the simplified two-dimensional unfolded Brillouin zone. The arrows show the nesting wavevectors $Q_1$ and $Q_2$. (d) Corbino geometry used to measure resistivity along the *c*-axis. The back surface of the crystal is identically wired to that which is illustrated for the front surface.

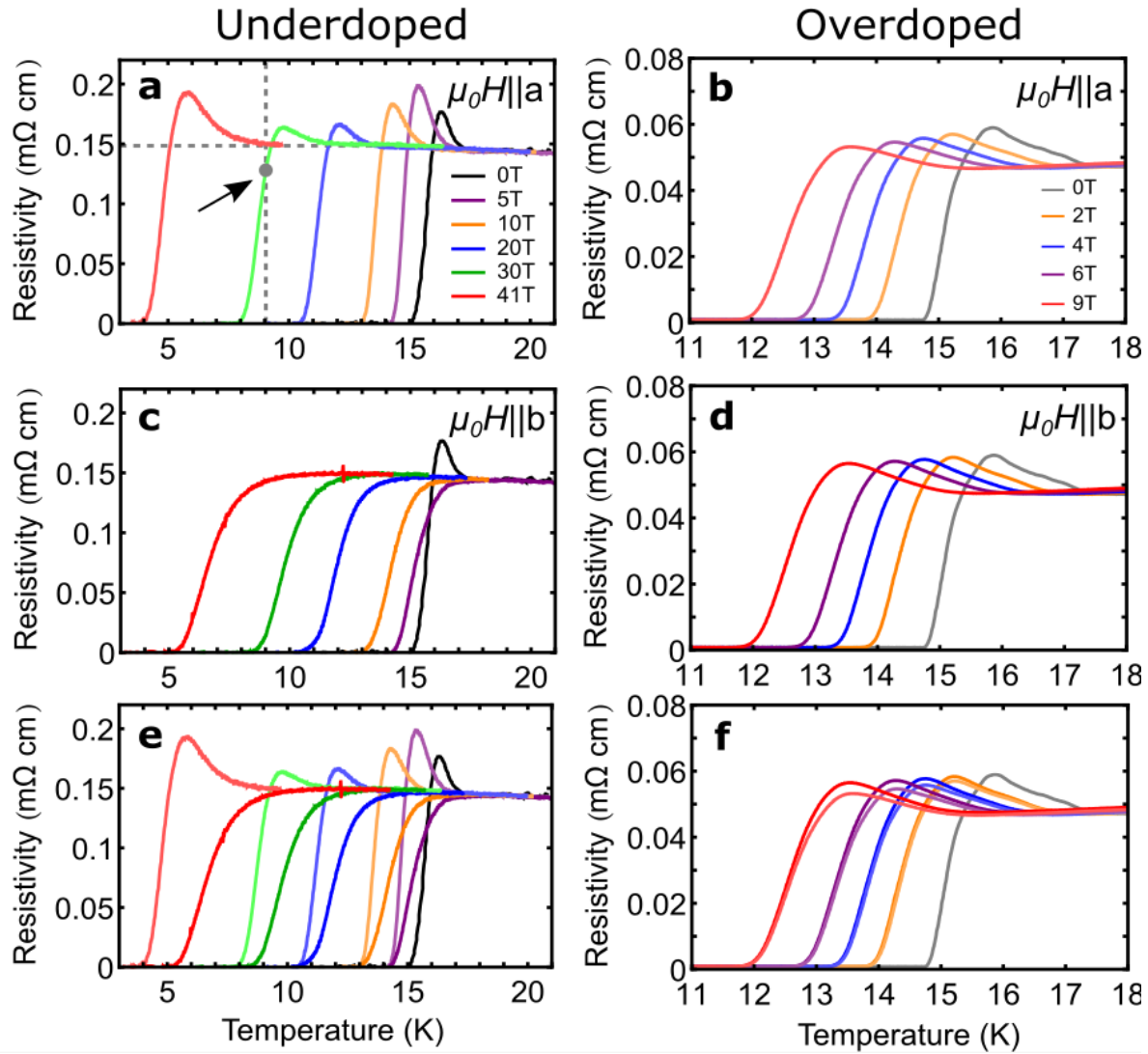

**Fig. 2** Resistivity measured using a Corbino geometry ($\rho_{zz}$) as a function of the temperature under varying field strengths applied along the *a*-axis for underdoped (a) and overdoped (b) samples. Data for field along the *b*-axis for underdoped and overdoped samples are given in (c) and (d) respectively. (e) Same data as in panels (a) and (c) but plotted together to illustrate the anisotropy in upper critical fields inherent to the underdoped sample. Similarly, (b) and (d) are combined in (f) to illustrate the contrasting lack of anisotropy. Treatment of the 90% criterion in the presence of the humps is illustrated in panel (a) with the arrow noting the point on the curve at which the resistivity drops to 90% of its normal state value. Notice that the peaks preceding the superconducting transition are not observed in the absence of strain, with their presence along a specific field crystallographic orientation requiring further theoretical understanding. Notice also, that at the highest magnetic fields the zero-resistance state (vortex solid phase) is achieved at higher temperatures for both crystallographic orientations, implying that the observed anisotropy is unrelated to the flux flow of vortices. The normal state resistivity of the 41T data in the underdoped sample was normalized to the normal state resistivity of the other curves.

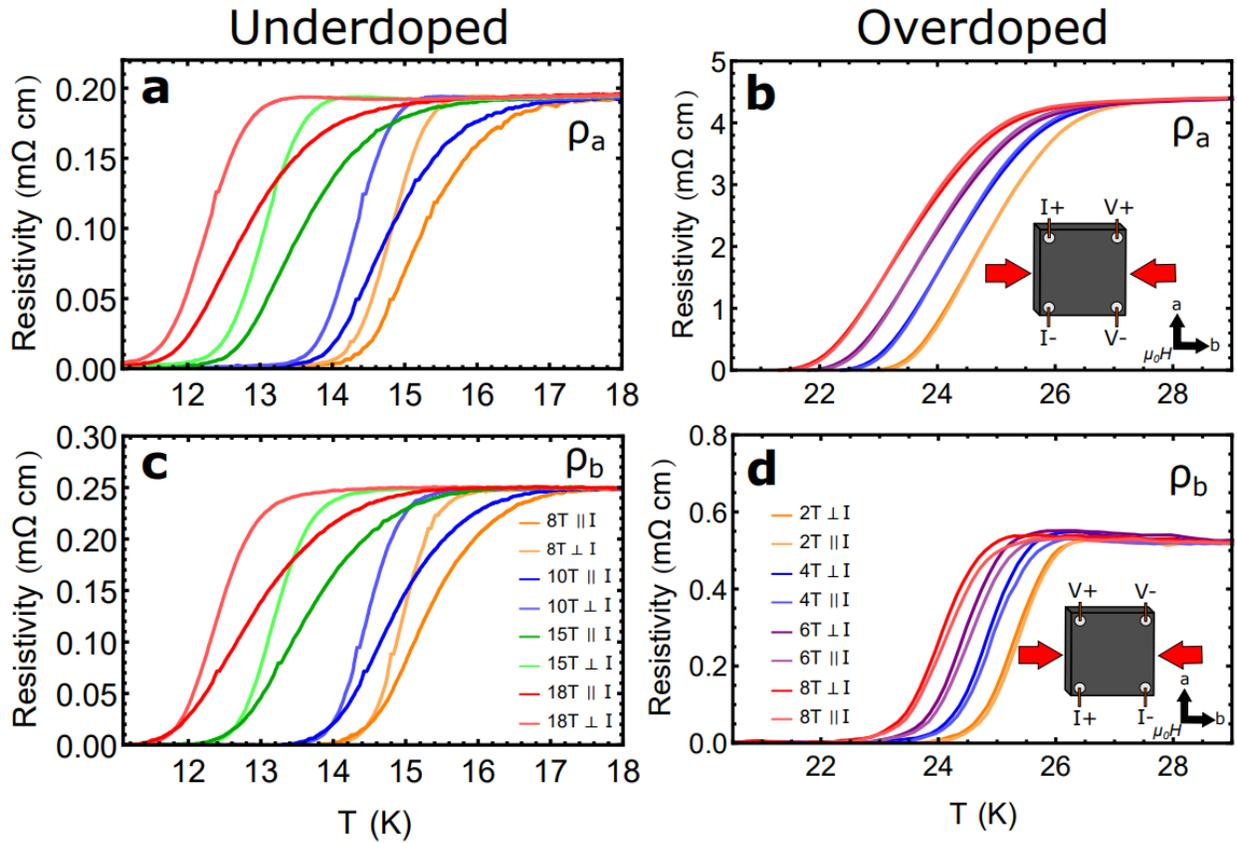

**Fig. 3** (a) and (b) Resistive transitions as a function of $T$ revealing the values of the upper critical fields $\mu_0 H_{c2}$ for strain applied along the $b$-axis and current along the $a$-axis and for both planar field orientations for underdoped and overdoped samples, respectively. Data for the currents along the $b$-axis is similarly presented for underdoped and overdoped samples in (c) and (d) respectively. Insets in (b) and (d) are schematics depicting the directions of the current (I+/-), strain (red arrows), and magnetic fields along crystalline axes $a$ and $b$ for each measurement. Regardless of the current orientation relative to strain, one always observes a larger inset field for currents aligned along the planar field.

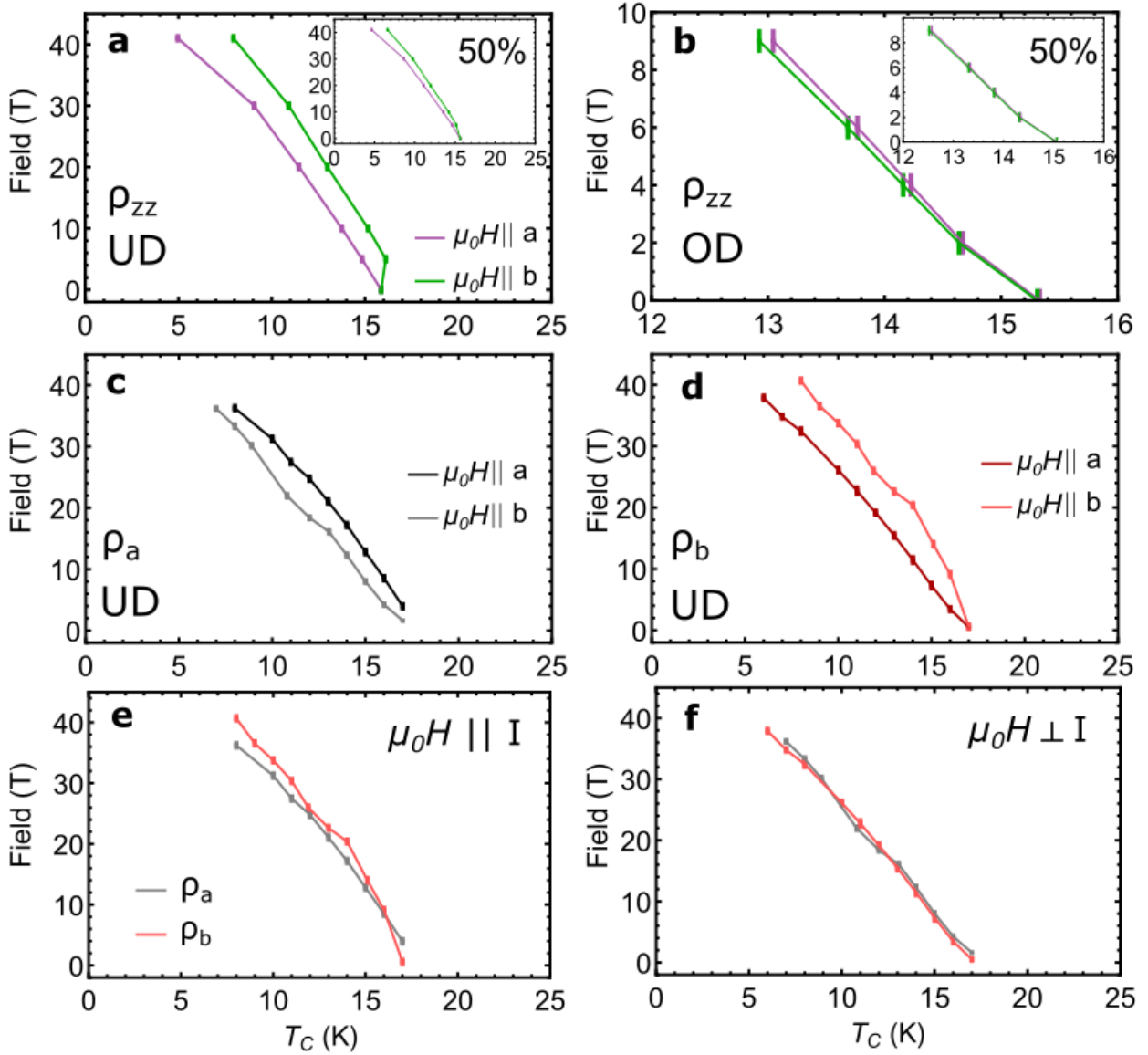

**Fig. 4** (a) $H_{c2}$ as a function of the temperature ($T$) when measured through the interplanar resistivity $\rho_{zz}$ for an underdoped sample using the 90% criterion. (b) $H_{c2}(T)$ for an overdoped sample as extracted from $\rho_{zz}$. The insets for (a) and (b) display the extracted $H_{c2}(T)$ using the 50% criterion. (c) and (d) $H_{c2}$ for an underdoped sample as a function of $T$ and as extracted from the in-plane resistivities, $\rho_a$ and $\rho_b$, respectively. (e) and (f) In-plane resistivity from (c) and (d) plotted such that the field and current are parallel and perpendicular respectively. Error bars are assigned to the extracted $H_{c2}$ to account for the humps. During the 90% criterion analysis, the normal state resistivity was considered from the peak of the hump to the flat region above the superconducting transition. The range of $H_{c2}$ resulting from the different definitions of the normal state resistivity is what determines the size of the error bars. This is a liberal estimation of the measurement error. Notice that the anisotropy in upper critical fields increases continuously as the temperature is lowered, in contrast to the anisotropy extracted for the activation energy for vortex flux flow (see Fig. S5). Therefore, the anisotropy in upper critical fields extracted using the 90% of the resistive transition criteria ought to be intrinsic, or is unrelated to vortex physics.

*email: balicas@magnet.fsu.edu
**email: pdai@rice.edu

Supplementary Material

To prepare single crystals of Ba(Fe$_{1-x}$M$_x$)$_2$As$_2$, the flux and Barium were mixed with a ratio of Ba:Fe$_{1-x}$M$_x$As = 1:4.5 in an Al$_2$O$_3$ crucible and sealed under vacuum in a quartz tube. The tube was first warmed up to 900°C over 20 hours, then to 1175°C for 10 hours, cooled to 1050°C in 25 hours, and lastly left to cool to room temperature with the furnace off. The underdoped samples were grown using Ni with x = 0.04, while the overdoped samples were grown using Co with x = 0.11. Larger crystals were aligned and later cut along the orthorhombic *a* and *b* axes into roughly 2x2 mm$^2$ squares using a Laue x-ray diffractometer and a wire saw respectively. Samples were then detwinned via uniaxial pressure with a small brass clamp [Fig. S3].

Two additional overdoped samples were measured nearer to the optimal doping to demonstrate no/negligible anisotropy in the overdoped regime. Figure S4(a) shows an analogous plot to Fig. 2 only with an overdoped sample with ~7% Co-content. Figures S4(b) and S4(c) are analogous plots to Fig. 3, but with nearly optimal doping ~6.5%. Figure S5 shows resistivity across superconducting transition. Figure S6 shows normal state field dependence resistivity. Both Figs. 5 and 6 are for Ba(Fe$_{1-x}$Ni$_x$)$_2$As$_2$ with x~0.04 sample. Figure S7 shows in-plane magnetic field dependence magnetoresistance in the normal state at 18 K, which is small and has no in-plane field directional dependence.

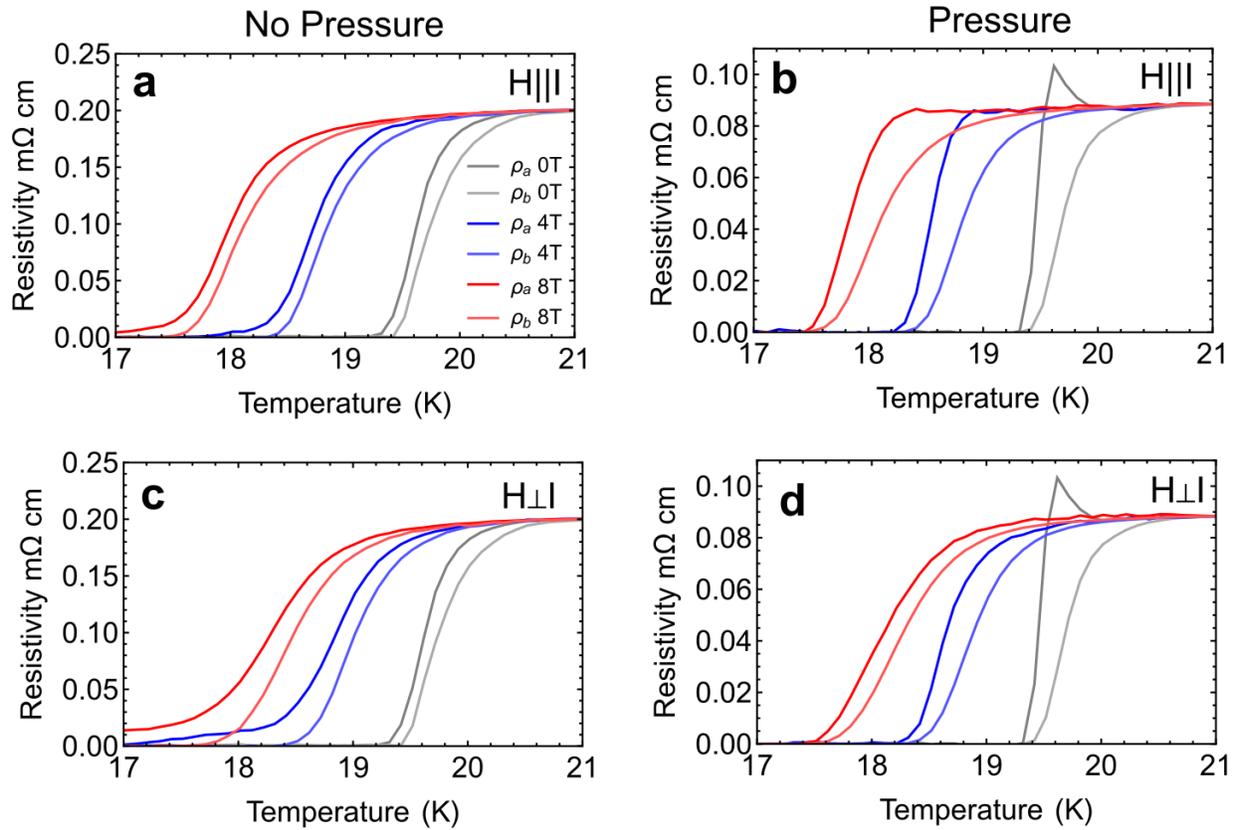

**Fig. S1**: (a) In-plane resistivity anisotropy with the field aligned along the current direction in the absence of strain. (b) With strain applied, the in-plane resistivity anisotropy with the field aligned along the current direction. (c) and (d) are the analogs to (a) and (b), respectively, with the field now aligned perpendicularly to the current.

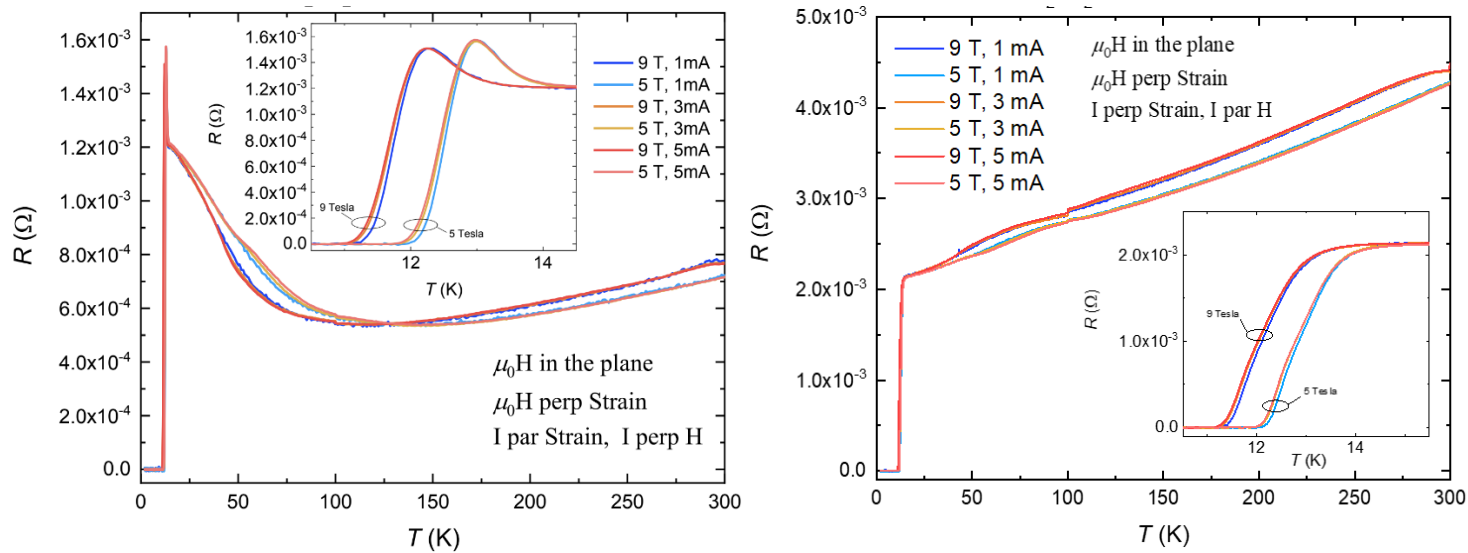

**Fig. S2**: Current dependence of the resistance of an underdoped sample showing resistivity along b (left) and a (right) axes. Three different currents (1, 3, & 5 mA) were measured at two fields (5,9 T). Inset highlights the temperature region immediately surrounding $T_c$.

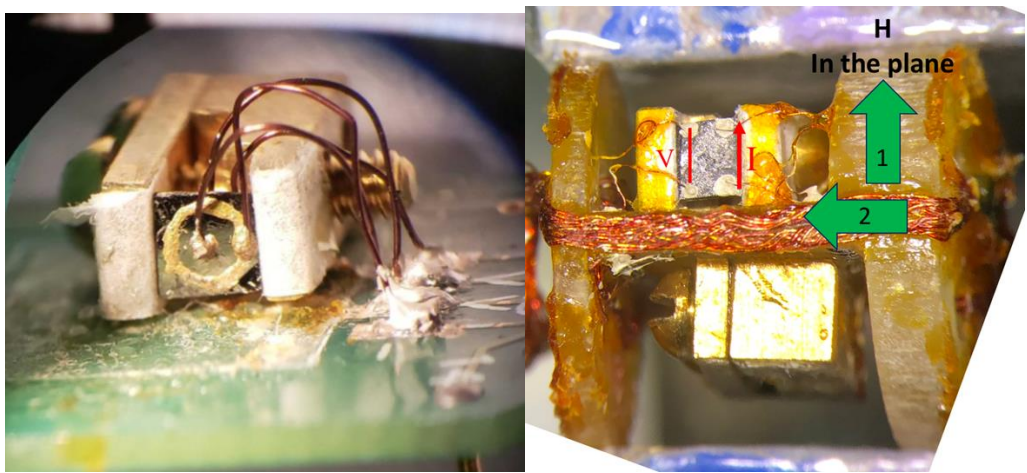

**Fig. S3**: Single crystal mounted on a brass clamp and wired in a Corbino geometry mounted on a Physical Property Measurements System puck (left). Two single crystals in two brass clamps mounted for measurements at the MagLab, with a diagram depicting voltage and current configuration for $\rho_a$ (right).

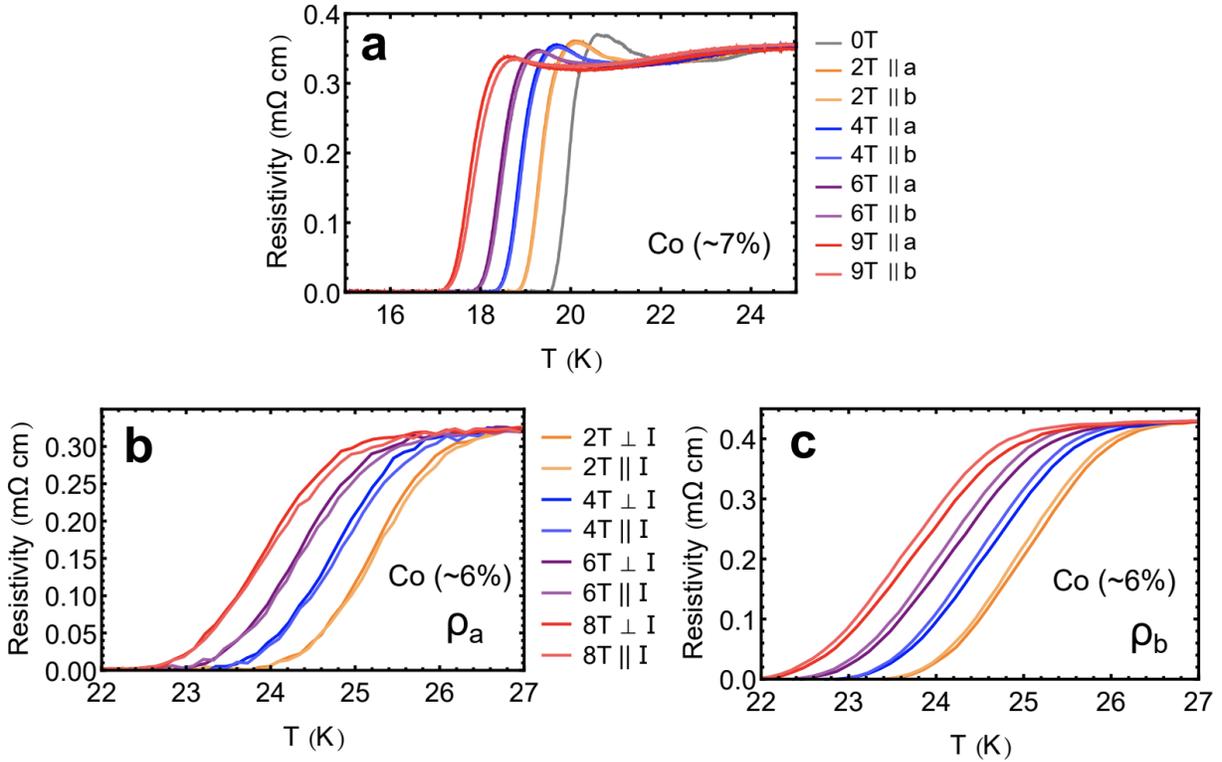

**Fig. S4**: (a) Overdoped sample measured with current along the *c*-axis using a Corbino geometry with roughly 7% Co-doping. (b,c) Resistivity $\rho_a$ and $\rho_b$, respectively, as a function of temperature for distinct orientations of the current relative to field for a different overdoped sample, close to optimally doped ~6% Co.

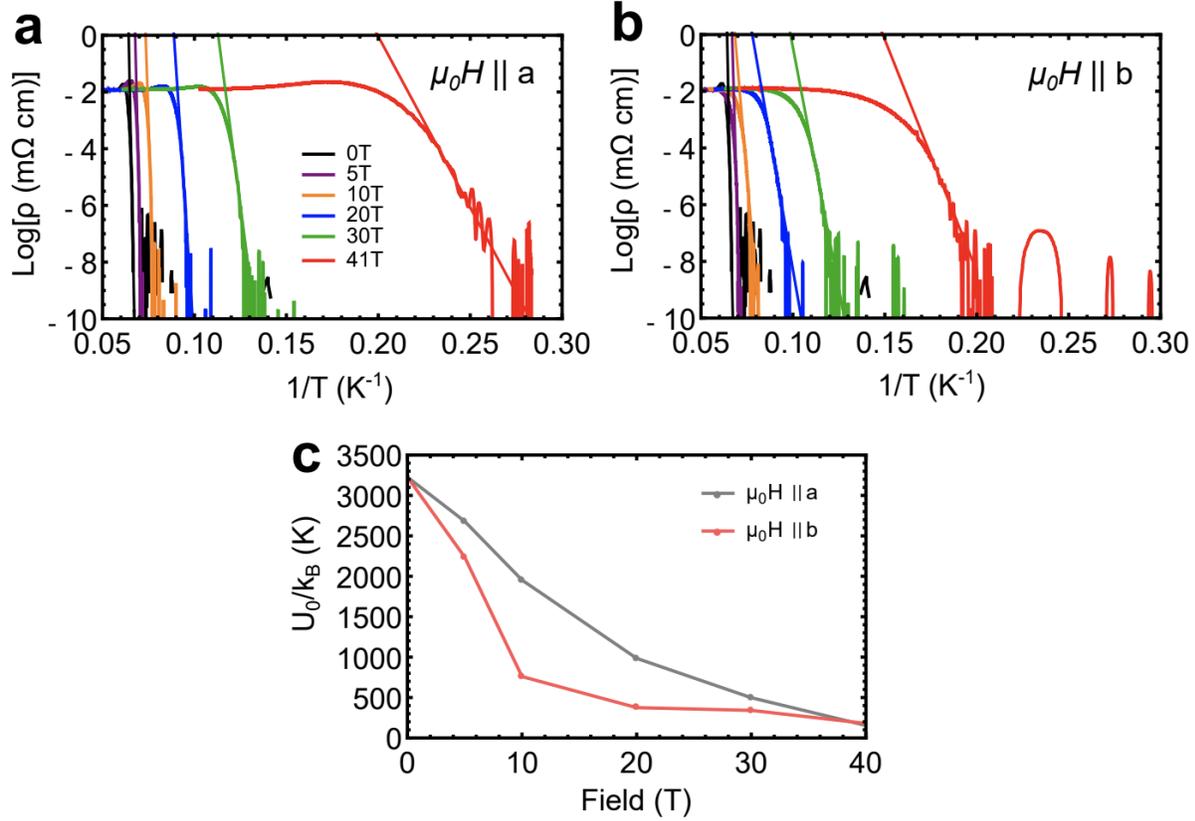

**Fig. S5**: (a) Resistivity across the superconducting transition in a logarithmic scale and for fields along the *a*-axis, as a function of the inverse temperature. Straight lines are linear fits from which we extract the activation energy $U_0$ for vortex flow. (b) Same as in (a) but for fields along the *b*-axis. (c) $U_0/k_B$ as a function of magnetic field for both field orientations.

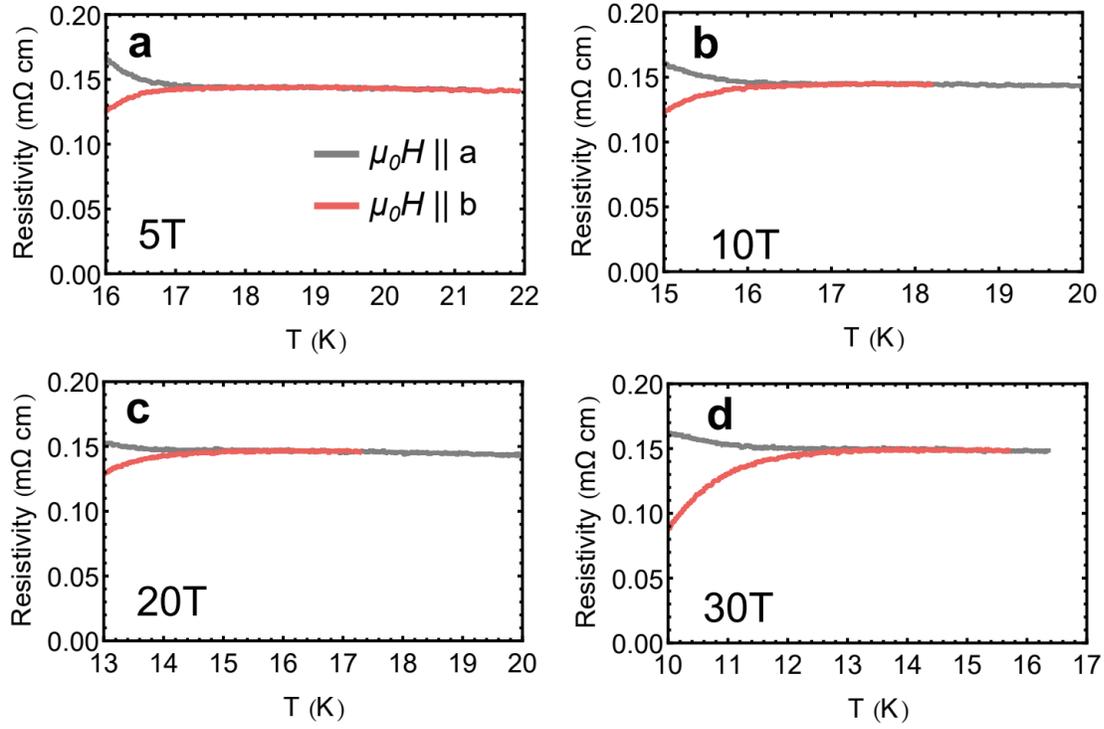

**Fig. S6**: (a-d) Field directional dependence of the normal state resistivity at 5, 10, 20, and 30 T. Data in this figure is the same as in Fig. 2(a,c,e) and is presented without scaling. This plot excludes data under $\mu_0 H = 41$ T since there is minimal overlap of the normal state resistivities.

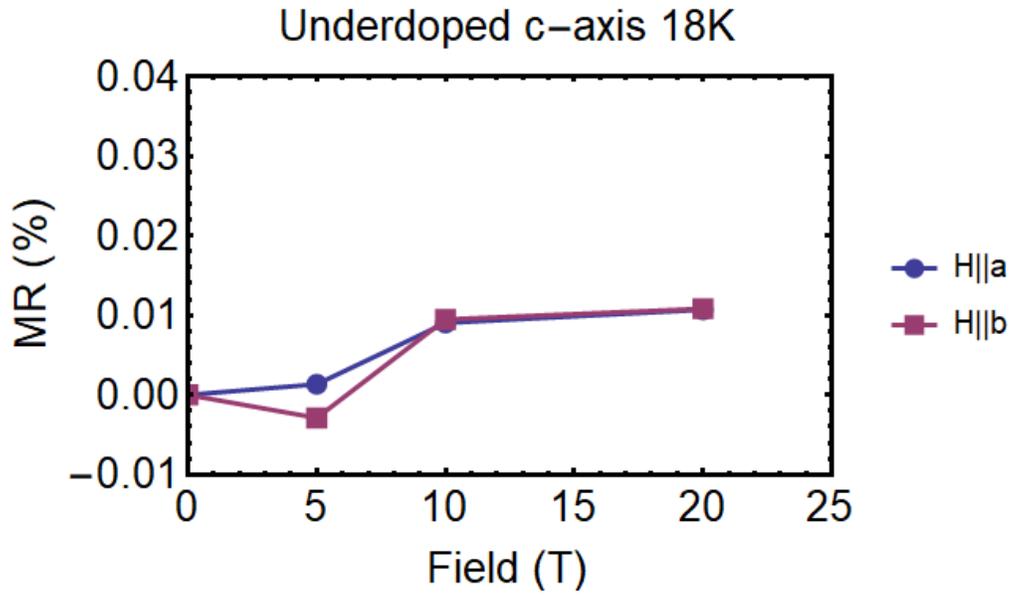

**Fig. S7**: Magnetoresistance (MR) of the underdoped crystal measured with *c*-axis resistivity and fields along the *a* and *b* axes from Fig. 2(a,c,e). The MR is negligible at 1% of 1% and the sample exhibits no in-plane directional field dependence.